\documentclass[elsarticle.cls,superscriptaddress, notitlepage]{revtex4-1}
\usepackage{amsmath,amssymb,amsthm}
\usepackage{braket}
\usepackage{times,txfonts}
\usepackage{braket}
\usepackage{color}
\usepackage{mathtools}
\usepackage{natbib}
\usepackage{latexsym}
\usepackage{tabularx, booktabs}
\usepackage{amssymb}
\usepackage{graphics,epstopdf}
\usepackage{graphicx}
\usepackage[colorlinks=true, citecolor=blue, urlcolor=blue ]{hyperref}
\usepackage{float}
\usepackage{graphicx}
\usepackage{amsfonts}
\usepackage{color}
\usepackage{caption}
\usepackage{subcaption}
\usepackage{wrapfig, framed, caption}

\begin{document}

\title{Quantum violation of macrorealism under multi-outcome two-parameter generalised measurements}

\author{Debarshi Das}
\email{dasdebarshi90@gmail.com}
\affiliation{Centre for Astroparticle Physics and Space Science (CAPSS), Bose Institute, Block EN, Sector V, Salt Lake, Kolkata 700 091, India}

\author{Arindam Gayen}
\email{arindam.gayen23@gmail.com}
\affiliation{Centre for Astroparticle Physics and Space Science (CAPSS), Bose Institute, Block EN, Sector V, Salt Lake, Kolkata 700 091, India}

\author{Ranit Das}
\email{ranitdas2@gmail.com}
\affiliation{Department of Physics, Indian Institute of Technology-Roorkee, Roorkee 247 667,  India}

\author{Shiladitya Mal}
\email{shiladitya.27@gmail.com}
\affiliation{Harish-Chandra Research Institute, HBNI, Chhatnag Road, Jhunsi, Allahabad 211 019, India}

\begin{abstract}
Generalised dichotomic quantum measurements are fully characterised by two real parameters,
dubbed as sharpness parameter and biasedness parameter. The trade-off between the degree of joint measurability, sharpness and biasedness of generalised measurements was known in the case of pairs of qubit observables. In the present work we generalise the notion of sharpness and biasedness measure of multi-outcome generalised measurements pertaining to multilevel systems. A trade-off between the amount of quantum mechanical (QM) violation of macrorealism (MR), sharpness and biasedness is established. Specifically we found that the minimum value of sharpness parameter, above which the QM violations of different necessary conditions of MR persist, decreases with increase in biasedness. We also analysed the effect of biasedness parameter on the magnitudes of QM violations of different necessary conditions of MR for multilevel spin systems. 
\end{abstract}

	\maketitle
	
\section{INTRODUCTION} \label{s1}
Description of nature in quantum physics is significantly different from that in classical physics. There are several no-go results revealing the departure of quantum theory from classical theory, which arise from statistics obtained by performing measurements on any system. Historically first no-go theorem is due to John Bell \cite{Bell} in response to Einstein-Podolsky-Rosen paradox \cite{EPR'1935} which states that local-realist theory cannot explain all the correlations obtained from local measurements on spatially separated entangled systems. Second no-go theorem, due to Kochen and Specker \cite{Kochen'1967}, states that there is no non-contextual hidden variable theory for a class of temporal correlations obtained from sequential commutative measurements on a single quantum system. The latest no-go theorem in this direction is due to Leggett and Garg in the form of an inequality, namely the Leggett-Garg inequality (LGI), which is a testable algebraic consequence of the notion of macrorealism (MR) \cite{LGI1,LGI2}. In recent years, investigations related to the LGI have been acquiring considerable significance, as evidenced by a wide range of studies; see, for example, a recent comprehensive review, Reference \cite{lgreview}.

It is to be noted here that all these no-go theorems are demonstrated through doing measurements on quantum systems. Measurements in quantum theory is another area where classical notion of measurement breaks down. In \cite{von'1935} von Neumann radically revised the theory of measurement in quantum mechanics. In this formulation quantum measurement is described by projective valued measure (PVM). Later this ideal notion of measurement, i.e., PVM is extended in order to incorporate realistic aspects of measurements, known as positive operator valued measure (POVM). The general description of quantum observables as POVMs gives rise to many operational possibilities not available within the set of standard observables (represented as PVM) \cite{unsharp2}. For example, POVM with the number of measurement outcomes being greater than the system dimension has application in unambiguous state discrimination problem \cite{usd1,usd2,usd3}. On the other hand, the imprecision or noise associated with measurement apparatus can be modeled by unsharp POVM \cite{unsharp2}, where the number of measurement outcomes is equal to the dimension of the system under consideration.

The impossibility of jointly measuring certain pairs of observables is an intriguing non-classical feature of quantum theory and another fundamental no-go theorem in the context of quantum measurement.  It follows that for two non-commuting observables to be jointly measurable, it is necessary that both of them are unsharp. Hence, there exists a trade-off between the degrees of non-commutativity and sharpness within the set of jointly measurable pairs of observables \cite{unsharp2,unsharp1,unsharp5}. Later the joint measurability of pairs of qubit observables is analysed to study a trade-off between the degrees of joint measurability, sharpness and another quantity called biasedness \cite{Stano,bias,bias2}. A two outcome experiment would be regarded as biased if one of the outcomes turned out to be preferred, whatever the preparation.

It was known that unsharpness of measurement reduces the possibility of observing quantum features. Apart from LGI, another two necessary conditions for testing MR, namely, Wigner's form of the Leggett-Garg inequality (WLGI) \cite{Saha} and no signalling in time (NSIT) condition \cite{Kofler} have been proposed. One important point to be stressed here is that the NSIT condition was previously proposed as quantum witness in \cite{qwitness1}. Furthermore, an upper bound for possible violations
of quantum witness equality due to quantum mechanics was derived and the quantum witness for precessing spin system was calculated \cite{qwitness2}. In the context of MR, POVMs has been introduced in the earlier works in association with emergence of classicality \cite{Kofler'2008,Saha,Mal-Majumdar,largelgi'2016}. In all these works it has been shown that the magnitudes of QM violations of all the three necessary conditions of MR decrease with decreasing precision of measurements \cite{Kofler'2008} or decreasing values of sharpness parameter, i.e., increasing unsharpness of the measurement \cite{Saha,Mal-Majumdar,largelgi'2016}. 

 In a recent work \cite{new} trade-off between sharpness and biasedness of dichotomic measurements has been explored by showing how biasedness of a measurement counters the effect of unsharpness in demonstrating violation of local realism and MR as well for relevant systems with minimal dimension. In another work \cite{pan} it is shown that there exists a particular biased POVM and initial single qubit state for which violation of all the necessary conditions of MR persists for any nonzero value of the sharpness parameter.
 
For probing violation of MR it is desired to consider systems with large quantum number like spin, large mass or large number of particles. Therefore it is required to extend the notion of sharpness and biasedness measure to arbitrary dimensional systems. This characterization is not entirely trivial due to ambiguity and the fact that the 2-dimensional case is too simple to reveal many relevant features.

In this work we generalise the notion of sharpness and biasedness measure of generalized multi-outcome measurements pertaining to multilevel systems. A trade-off between sharpness and biasedness of measurement is also investigated in the context of quantum mechanical (QM) violation of MR invoking a particular measurement scheme which is introduced in  \cite{bdroni} and it was shown that with such a measurement scheme there exists an initial state for which LGI, WLGI and NSIT can be violated upto its algebraic maximum for asymptotically large value of spin  \cite{bdroni,largelgi'2016}. In particular, here it is shown that the minimum value of the sharpness parameter ($\lambda_{th}$) over which QM violation of any necessary condition of MR persists, decreases with increasing spin value ($j$) of the system for any fixed value of the biasedness parameter, $\gamma$. For any fixed value of $j$, $\lambda_{th}$ decreases with increasing biasedness; and for any fixed values of $j$ and $\lambda$, the magnitude of QM violation of any necessary condition of MR increases with increase in $\gamma$. It is also found that in the asymptotic limit of spin value, our proposed biased unsharp measurement reduces to unsharp measurement with zero biasedness. 
 
In the following Section, we explain in a nutshell the various necesesary conditions proposed for testing MR. In Section \ref{s3}, the notion of POVM and the generalisation of biased unsharp measurement for multi-outcome spin measurement have been discussed. In Section \ref{s4}, the relevant features of the system and measurement scheme is presented. In Section \ref{s5}, the key results obtained using LGI, WLGI and NSIT are demonstrated. Finally we conclude in Section \ref{s6} mentioning directions for future studies.

\section{Various necessary conditions of MR} \label{s2}
The concept of MR is codified by the following two assumptions: {\it 1) Macroscopic realism per se:} At any instant, a macroscopic system is in any one of the available definite states, regardless of any measurement, such that all its observable properties have definite values. {\it 2) Noninvasive measurability (NIM):} It is possible, in principle, to determine which of the states the system is in, without influencing the state itself or the system's evolution.\\
QM violation of any of the following necessary conditions of MR implies the negation of the concept realism or NIM or both in the underlying hidden variable theory reproducing all QM predictions.

\textbf{Leggett-Garg inequality (LGI)}: LGI \cite{LGI1,LGI2} is derived as a testable algebraic consequence of the  deterministic form of MR. Let us consider temporal evolution for a two state system where the available states are, say, $1$ and $2$. Let $Q(t)$ be an observable quantity such that, whenever measured, it is found to take a value $+1 (-1)$ depending on whether the system is in $1 (2)$. Next, consider a collection of sets of experimental runs, each set of runs starting from the identical initial state such that on the first set of runs $Q$ is measured at times $t_{1}$ and $t_{2}$, on the second at $t_{2}$ and $t_{3}$ and on the third at $t_{1}$ and $t_{3}$ $(t_{1} < t_{2} <t_{3})$. Using the deterministic consequence of the assumptions of realism and NIM, one obtains
\begin{equation}
	\label{lgi}
	K_{LGI}= C_{12} +C_{23} - C_{13} \leq 1,
\end{equation}
where $C_{ij}$ = $\langle Q_i Q_j\rangle$ is the two time correlation function of the variable $Q$ measured at times $t_i$ and $t_j$. Left hand side of the inequality (\ref{lgi}) is an experimentally measurable quantity. This is the LGI imposing non-invasive realist constraint on the temporal correlations pertaining to any two level system. The magnitude of the QM violation of LGI is quantified by the positive value of $(K_{LGI} - 1)$.

\textbf{Wigner's form of the LGI (WLGI)}: In \cite{Saha}, WLGI is derived as a testable algebraic consequence of the probabilistic form of MR. Here the notion of realism implies the existence of overall joint probabilities $\rho (Q_1, Q_2, Q_3)$ pertaining to different combinations of definite values of observables or outcomes for the relevant measurements, while the assumption of NIM implies that the probabilities of such outcomes would be unaffected by measurements. Hence, by appropriate marginalization, the observable probabilities can be obtained. For example, the observable joint probability $P(Q_{2}+, Q_{3}-)$ of obtaining the outcomes $+1$ and $-1$ for the sequential measurements of Q at the instants $t_2$ and $t_3$, respectively, can be written as
\begin{align}
P(Q_{2}+, Q_{3}-) &= \sum_{Q_{1}=\pm 1} \rho(Q_{1}, +, -) = \rho (+, +, -) + \rho(-, +, -).
\end{align}
Expressing similarly the other measurable marginal joint probabilities $P(Q_{1}-, Q_{3}-)$ and
$P(Q_{1}+, Q_{2}+)$, we get 
\begin{align}
&P(Q_{1}+, Q_{2}+) + P(Q_{1}-, Q_{3}-) - P(Q_{2}+, Q_{3}-) = \rho (+, +, +) + \rho(-, -, -).
\end{align}
Then invoking non-negativity of the overall joint probabilities occurring on the right hand side of the above equation, the following form of WLGI is obtained in terms of two-time joint probabilities
\begin{align}
\label{eqq}
K_{WLGI} &= P(Q_{2}+, Q_{3}+) - P(Q_{1}-, Q_{2}+) - P(Q_{1}+, Q_{3}+)  \leq 0.
\end{align}
Similarly, other forms of WLGI involving any number of two-time joint probabilities can be derived by using various combinations of the observable joint probabilities. The magnitude of the QM violation of WLGI is quantified by the positive value of $K_{WLGI}$.

\textbf{The condition of no signalling in time (NSIT)}: NSIT condition states that the measurement outcome statistics for any observable at any instant is \textit{independent} of whether any prior measurement has been performed \cite{Kofler}. Let us consider a system whose time evolution occurs between two possible states. Probability of obtaining the outcome $+1$ for the measurement of a dichotomic observable $Q$ at an instant, say, $t_3$ \textit{without} any earlier measurement being performed, is denoted by $P(Q_3 = +1)$. NSIT requires that $P(Q_3 = +1)$ should remain \textit{unchanged} even when an earlier measurement is made at $t_2$. Mathematically NSIT can be expressed as an equation given by,
\begin{align}
	K_{NSIT} &= P(Q_3 = -1) - [P(Q_2 = +1, Q_3 = -1) + P(Q_2 = -1, Q_3 = -1)] = 0.
	\label{eq3}
\end{align}
The magnitude of the QM violation of NSIT is quantified by the nonzero value of $K_{NSIT}$.

	\section{Modelling biased unsharp POVM in case of multiple outcome spin measurements} \label{s3}
	
Projective valued measurement (PVM) is a set of projectors that add to identity, i.e., $A\equiv \lbrace P_{i}\vert\sum P_{i}=\mathbb{I}, P_i^2 = P_i \rbrace$ (where  the Hermitian operators $P_i$s are projectors). The probability of getting the $i$-th outcome is given by,  Tr$[ \rho P_i ]$ for the state $\rho$ (here $\rho$ is a Hermitian, positive operator and Tr$[\rho] = 1$). On the other hand, positive operator valued measurement (POVM) is a set of positive operators that add to identity, i.e., $E\equiv \lbrace E_{i}\vert\sum E_{i}=\mathbb{I},0< E_i\leq \mathbb{I}\rbrace$. The probability of getting $i$-th outcome is Tr$[\rho E_i]$.
 
Unbiased unsharp measurement \cite{unsharp2} is a particular example of POVM, which is characterised by sharpness parameter ($\lambda$). The corresponding effect operators associated with dichotomic unbiased unsharp measurement are given by,
	\begin{eqnarray}
	E^{\pm} =\lambda P^{\pm}+(1-\lambda)\frac{\mathbb{I}}{2},
	\label{undi}
	\end{eqnarray}
where $P^{\pm}$ are sharp projectors. For $E^{\pm}$ to be valid effect operator, the positivity ($E^{\pm} \geq 0$) and normalisation ($E^+ + E^- = \mathbb{I}$) conditions have to be satisfied. From these conditions we get $-1 \leq \lambda \leq 1$. Note that $(1 - \lambda)$ characterises the amount of unsharpness associated with the measurement. On the other hand, `$\lambda$' characterises the `closeness' of a measurement to projective measurement and $\lambda =1$ implies that the measurement is projective. Hence, negative values of $\lambda$ have no physical significance. Furthermore, the characteristics of the mathematical results obtained in our context does not change if one uses negative values of $\lambda$. Therefore, we assume that $0 \leq \lambda \leq 1$ without loss of generality. It is to be noted that $\lambda =0$ implies trivial effect operators. In other words, $\lambda =0$ implies that the effect operators are nothing but identity operators which means no measurement takes place. Hence, we consider $0 < \lambda \leq 1$ for physically meaningful unbiased unsharp measurements. Unbiased unsharp measurement is not the most general dichotomic POVM.

The most general dichotomic POVM is biased unsharp measurement, which is characterized by two real parameters- sharpness parameter ($\lambda$) and biasedness parameter ($\gamma$) \cite{Stano,bias,bias2}. The corresponding effect operators are given by,
	\begin{eqnarray}
	E^{\pm} =\lambda P^{\pm}+(1\pm\gamma-\lambda)\frac{\mathbb{I}}{2}.
	\label{biundi}
	\end{eqnarray}
From positivity and normalization conditions we get $|\gamma| +|\lambda |\leq 1$. As mentioned earlier, we will consider $\lambda > 0$ from physical point of view. Moreover, the value of $\gamma$ characterises the `amount' of  biasedness (a particular form of nonidealness) associated with the measurements. Hence, without any loss of generality, we also consider $\gamma \geq 0$ as `negative amount nonidealness' has no physical meaning. Note that negative $\gamma$ does not alter the significance of the results obtained. Equation (\ref{biundi}) with $\gamma = 0$ gives the aforementioned unbiased unsharp measurement.

Now consider sharp spin-$z$ component ($J_z$) measurement on a multilevel ($2j+1$ dimensional) spin-$j$ system, where the possible outcomes are the eigenvalues of $J_z$ operator and are denoted by $m$ (the possible values of $m$ are $-j$, $-j+1$, $-j+2$, ..., $j-1$, $j$). The projectors onto the state which is an eigenstate of the sharp $J_z$ observable with eigenvalue $m$  is denoted by $P^m$.  Here we generalize the effect operators corresponding to the biased unsharp POVM in the following way
\begin{equation}
F^{m}=\lambda P^{m}+(1+m\gamma -\lambda) \frac{\mathbb{I}}{2j+1}.
\label{newbiun1}
\end{equation}
In this case also, we consider that $\lambda > 0$ and $\gamma \geq 0$ from physical point of view as described earlier. Now we  check the conditions of positivity and normalization for these effect operators to form valid POVM.

\textbf{Completeness:} This condition is satisfied as all the effects sum to identity.
\begin{eqnarray}
		\sum_{m=-j}^{j}F^{m}=\sum_{m} \big[ \lambda P^{m}+(1+m\gamma -\lambda) \frac{\mathbb{I}}{2j+1} \big]
		=\mathbb{I}.
	\end{eqnarray}
	
\textbf{Positivity:} From the requirement of positivity of effects we get,
\begin{equation}
1+2j\lambda +m\gamma\geq 0,
\end{equation}
and 
\begin{equation}
1+m\gamma-\lambda\geq 0.
\end{equation}
These two inequalities imply that
\begin{equation}
-\frac{1+m\gamma}{2j}\leq\lambda\leq 1+m\gamma.
\end{equation}
Here the lower bound and the upper bound are `$m$' dependent.  Now, we want `$m$' independent lower bound and upper bound of $\lambda$. Since we have considered that $\gamma \geq 0$, we can take 
	\begin{eqnarray}
		\label{eq}
		-\frac{1-j\gamma}{2j}\leq \lambda\leq 1-j\gamma.
	\end{eqnarray}
As we have already assumed that $\lambda > 0$, we can consider the permissible range of $\lambda$ as
	\begin{eqnarray}
	\label{lambda}
		0 < \lambda\leq1-j\gamma.
	\end{eqnarray}
From the above mentioned range of $\lambda$, we get an upper bound for $\gamma$ also. This is given by, 
	\begin{eqnarray}
	\label{gamma}
     0 \leq \gamma < \frac{1}{j}.
	\end{eqnarray}
Hence, for multilevel systems we obtained the form of effect operators and the permissible ranges for $\lambda$ and $\gamma$ are given by inequalities (\ref{lambda}) and (\ref{gamma}) respectively.

Given the above specification of the effect operators, the probability of an outcome, say $m$, is given by Tr$(\rho F^m)$ for which the post-measurement state is given by L{\"u}der transformation rule, $(\sqrt{F^m} \rho \sqrt{F^m}^{\dagger}) /$ Tr$(\rho F^m)$, $\rho$ being the state of the system ($\rho$ is a Hermitian, positive operator and Tr$[\rho] = 1$) on which measurement is done.  Note that from Equation (\ref{newbiun1}), we can write
\begin{align}
F^{m} &= \lambda P^{m}+\frac{1+m\gamma -\lambda}{2j+1} \sum_{k=-j}^{j} P^{k} = \frac{1+ 2 j \lambda + m \gamma}{2j+1} P^{m} + \frac{1+m\gamma -\lambda}{2j+1} \sum_{\substack{k= -j \\ k \neq m}}^{j} P^{k}.
\label{newbiun2}
\end{align}
Hence, from Equation (\ref{newbiun2}) we get
\begin{align}
\sqrt{F^m} &= \Bigg(\sqrt{\frac{1+ 2 j \lambda + m \gamma }{2j+1}} \Bigg) P^{m}  +\Bigg(\sqrt{\frac{1+m\gamma -\lambda}{2j+1}} \Bigg)\sum_{\substack{k= -j \\ k \neq m}}^{j} P^{k} \nonumber \\
&= \Bigg(\sqrt{\frac{1+ 2 j \lambda + m \gamma}{2j+1}} - \sqrt{\frac{1+m\gamma -\lambda}{2j+1}} \Bigg) P^{m}  + \Bigg(\sqrt{\frac{1+m\gamma -\lambda}{2j+1}} \Bigg) \mathbb{I}.
\end{align}

Here it should be noted that while for qubits ``biased unsharp measurement" \cite{Stano,bias,bias2} is the most general kind of two outcome POVM, for multi-outcome measurements these two parameters do not constitute the most general unique form of POVM but a reasonable one.

	\begin{figure}[!t]
			\resizebox{8.5cm}{7.5cm}{\includegraphics{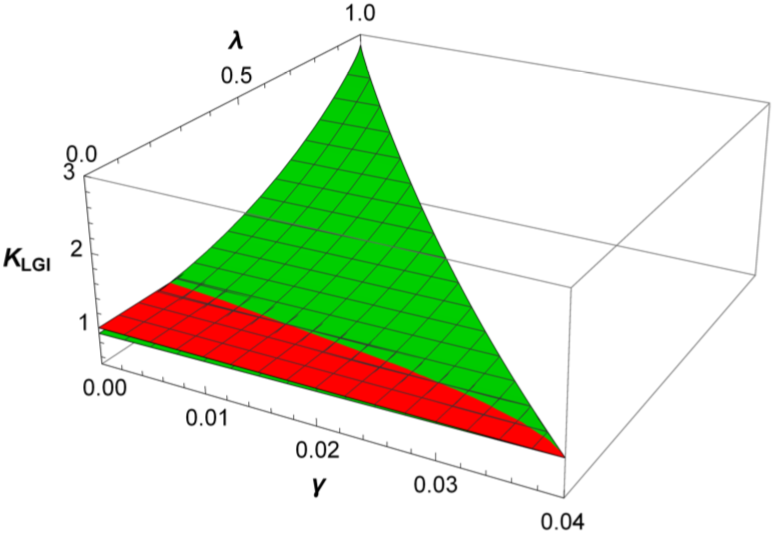}}
			\caption{\footnotesize The green surface denotes $K_{LGI}$ for different values of the sharpness parameter $\lambda$ and biasedness parameter $\gamma$ (satisfying $ 0 \leq \gamma < \frac{1}{j}$ and $0 < \lambda\leq1-j\gamma$) when the spin of the system $j=25$. The red surface denotes $K_{LGI} = 1$. $K_{LGI} > 1$ implies QM violation of LGI. In other words, QM violation of LGI occurs when the green surface lies above the red surface.} \label{fig1N}
		\end{figure} 

    \section{Setting up the measurement scenario} \label{s4}
Now we describe briefly the projective measurement scheme introduced in \cite{bdroni}, which is treated with sharpness and biasedness parameter in our calculations. Consider a system of spin $j$ in an external uniform magnetic field of magnitude $B_0$ along the $x$ direction. The relevant Hamiltonian is $(\hbar = 1)$ given by,
\begin{equation}
	\label{hh}
	H = \Omega J_x,
\end{equation}
where $\Omega$ is the angular precession frequency $(\propto B_0)$ and $J_x$ is the $x$ component of spin angular momentum.

Consider measurements of the $z$-component of spin. We define the observable $Q$ in such a way that $Q=-1$ when $m=-j$ and for any other values of $m$ ranging from $-j +1$ to $+j$, $Q = +1$, where $m$ is the output of $z$-component of spin measurement.

The above grouping scheme (call it one versus remaining) is chosen as it leads to maximal possible violation of all the necessary conditions of MR \cite{largelgi'2016}. We have not considered arbitrary grouping scheme (general case) as it reduces the magnitudes of the QM violations of different necessary conditions of MR \cite{largelgi'2016}. As our aim is to obtain the trade-off between unshapness and biasedness in the context of MR, it is best illustrated with the one versus remaining measurement scheme. In the following we use unsharp-biased version of this projective measurement scheme.

We initialize the system so that at $t$=$0$, the system is in the state $|-j;j\rangle$; where $|m;j\rangle$ denotes the Eigenstate of the $J_z$ operator with Eigenvalue $m$. We consider measurements of $Q$ at three successive times $t_1$, $t_2$ and  $t_3$ $(t_1<t_2<t_3)$ \& set the measurement times  as $\Omega t_1 = \Pi $ and $\Omega(t_2 - t_1)$ = $\Omega(t_3 - t_2)$  = $\frac{\Pi}{2}$.

One important point to be stressed here is that the above choices of measurement time intervals may not give the maximum quantum violation of the LGI, WLGI, or NSIT for arbitrary values of $j$ in the context of our measurement scheme. However, this choice suffices to give a representative indication of the trade-off between unshapness and biasedness associated with the measurements in the context of QM violations of MR. These particular choice of measurement time intervals enables us to carry out all the calculations analytically with the help of Wigner's D-matrix formalism.

		{\centering
		\begin{figure}[!t]
			\resizebox{8.5cm}{7.5cm}{\includegraphics{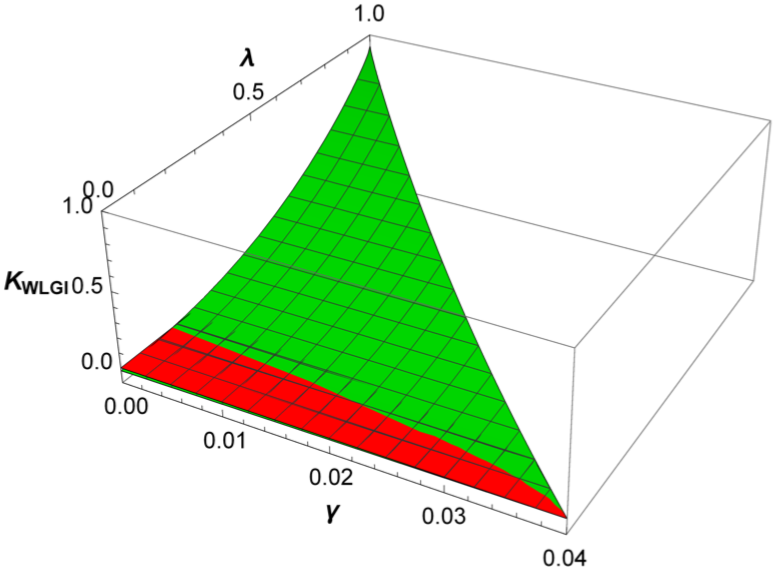}}
			\caption{\footnotesize The green surface denotes $K_{WLGI}$ for different values of the sharpness parameter $\lambda$ and biasedness parameter $\gamma$ (satisfying $ 0 \leq \gamma < \frac{1}{j}$ and $0 < \lambda\leq1-j\gamma$) when the spin of the system $j=25$. The red surface denotes $K_{WLGI} = 0$. $K_{WLGI} > 0$ implies QM violation of WLGI. In other words, QM violation of WLGI occurs when the green surface lies above the red surface.} \label{fig2N}
		\end{figure} 
		}
		
{\centering		
		\begin{figure}[!t]
			\resizebox{8.5cm}{7.5cm}{\includegraphics{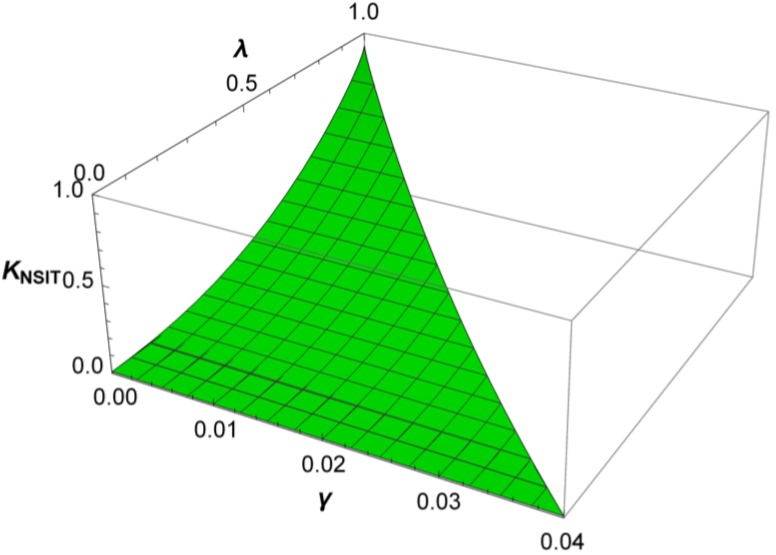}}
			\caption{\footnotesize The green surface denotes $K_{NSIT}$ for different values of the sharpness parameter $\lambda$ and biasedness parameter $\gamma$ (satisfying $ 0 \leq \gamma < \frac{1}{j}$ and $0 < \lambda\leq1-j\gamma$) when the spin of the system $j=25$. The red surface denotes $K_{NSIT} = 0$. $K_{NSIT} \neq 0$ implies QM violation of NSIT. In other words, QM violation of NSIT occurs when the green surface lies above or below the red surface. In the present case, the red surface is not visible as the green surface always lies above the red surface, i.e., $K_{NSIT} > 0$ for any values of $\lambda$ and $\gamma$.} \label{fig3N}
		\end{figure} 
		}

	\section{Analysis with LGI, WLGI and NSIT} \label{s5}
In order to calculate the expectation values and joint probabilities appearing in the aforementioned forms of LGI, WLGI and NSIT, we proceed by expressing the time evolution operator from the initial time $t=0$ to the instant of first measurement $t=t_1$ by $U(t_1 - 0) = \exp{[-i\pi J_x]}$,  and all the subsequent measurements are equispaced in time. Using the Wigner's D matrix formalism the expression of a typical joint probability distribution is obtained of the form given by,
 
 \begin{align}
 	P(Q_1 +, Q_2 -)&= \sum_{k=-j+1}^{j} P(m_1 = k, m_2 =-j) \nonumber\\
 	&= \sum_{k=-j+1}^{j} Tr[ F_{-j} U_{\Delta t_{2}} \sqrt{F_k} U_{\Delta t_{1}} \rho_i U_{\Delta t_{1}}^\dagger \sqrt{F_k}^\dagger U_{\Delta t_{2}}^\dagger] \nonumber\\
 &=	- \frac{(-2 + 2 \lambda - \gamma) j (1 - \lambda - j \gamma)}{(1 + 2 j)^2}  + \frac{2^{-2 j} (-1 + 2^{2 j})\lambda (1 - \lambda - j \gamma)}{(1 + 2 j)},
 	\label{ee}
 \end{align}
 where $P(m_1 = k, m_2 =-j)$ denotes the joint probability of obtaining the outcomes $k$ ($k$ $\in$ $\{-j+1, -j+2, -j+3, ...,j \}$) and $-j$ when the biased unsharp measurements of the $z$-component of spin observable are performed at instances $t_1$ and $t_2$ respectively. The initial state of the system at $t=0$ is denoted by $\rho_i$ = $|-j;j\rangle \langle -j; j|$. The time evolution operator from $t=0$ to $t=t_1$ is denoted by $U_{\Delta t_{1}}$ = $U(t_1 - 0)$ = $\exp{[-i\pi J_x]}$. The time evolution operator from $t=t_1$ to $t=t_2$ is denoted by $U_{\Delta t_{2}}$ = $U(t_2 - t_1)$ = $\exp{[-i\frac{\pi}{2} J_x]}$.

 Using such expressions for the joint probabilities, we analytically obtain the expressions of $K_{LGI}$, $K_{WLGI}$ and $K_{NSIT}$.

For any value of spin $j$ and any value of the biasedness parameter $\gamma$ we observe the following features: the magnitudes of QM violations of LGI, WLGI or NSIT decreases with decreasing values of the sharpness parameter $\lambda$. Below a certain value of $\lambda$ QM violation of LGI (or WLGI) disappears. However, QM violation of NSIT always persists for any value of $j$ and any value of $\gamma$ and $\lambda > 0$. These features are illustrated in Figures \ref{fig1N}, \ref{fig2N} and \ref{fig3N} where we have plotted  $K_{LGI}$, $K_{WLGI}$ and $K_{NSIT}$ for different values of $\lambda$ and $\gamma$ (satisfying $ 0 \leq \gamma < \frac{1}{j}$ and $0 < \lambda\leq1-j\gamma$) for $j=25$. For other values of spin $j$, we observe that the plots of $K_{LGI}$, $K_{WLGI}$ and $K_{NSIT}$ versus $\lambda$ and $\gamma$ are of similar nature.

{\centering
 	\begin{table}
 		\begin{tabular}{|*{5}{c|}}
 			\hline
 			\textit{\textbf{\textbf{}}} & \textit{\textbf{\textbf{}}} & \multicolumn{3}{|c|}{\textit{\textbf{Threshold Sharpness for}}}\\
 			\textit{\textbf{\textbf{Spin ($j$)}}} & \textit{\textbf{\textbf{Value}}} & \multicolumn{3}{|c|}{\textit{\textbf{}}}\\
 			\cline{3-5} 
 			\textit{\textbf{\textbf{of the}}} & \textit{\textbf{\textbf{of}}} & \textit{\textbf{LGI}} & \textit{\textbf{WLGI}} & \textit{\textbf{NSIT}} \\
 			\textit{\textbf{\textbf{system}}} & \textit{\textbf{\textbf{$\gamma$}}} & \textit{\textbf{}} & \textit{\textbf{}} & \textit{\textbf{}} \\
 			\textit{\textbf{\textbf{}}} &  \textit{\textbf{\textbf{}}} & \textit{\textbf{($\lambda_{th}^{LGI}$)}} & \textit{\textbf{($\lambda_{th}^{WLGI}$)}} & \textit{\textbf{($\lambda_{th}^{NSIT}$)}} \\
 			\hline
 			\hline
 			& $0$ & $0.30$ & $0.23$ & $0$ \\
 			$15$ & $0.030$ & $0.22$ & $0.16$ & $0$ \\ 
 			& $0.050$ & $0.14$ & $0.12$ & $0$ \\
 			\hline
 			& $0$ & $0.26$ & $0.20$ & $0$ \\
 			$20$ & $0.025$ & $0.18$ & $0.14$ & $0$ \\ 
 			& $0.040$ & $0.10$ & $0.09$ & $0$ \\
 			\hline
 			& $0$ & $0.23$ & $0.18$ & $0$ \\
 			$25$ & $0.020$ & $0.16$ & $0.13$ & $0$ \\ 
 			& $0.030$ & $0.11$ & $0.09$ & $0$ \\
 			\hline
 		\end{tabular}
 		\caption{For a fixed value of spin ($j$), threshold sharpness of LGI ($\lambda_{th}^{LGI}$) and that for WLGI ($\lambda_{th}^{WLGI}$) decrease with increasing values of the biasedness parameter ($\gamma$) associated with the measurements; on the other hand, the threshold sharpness for NSIT ($\lambda_{th}^{NSIT}$) is zero for all values of the biasedness parameter ($\gamma$) for any value of spin ($j$).} \label{tab1}
 	\end{table}
}

The minimum value of the sharpness parameter $\lambda$ for which the QM violations of LGI, WLGI or NSIT just begin to disappear for a given biasedness $\gamma$ of the measurements for a given spin $j$ system is called the threshold sharpness $\lambda_{th}$ for that value of $\gamma$ and $j$ pertaining to the measurement.

The range of the sharpness parameter $\lambda$ for which the QM violation of LGI persists in a spin `$j$' system for a given biasedness $\gamma$ of the measurement is given by $(\lambda_{th}^{LGI}$, $1-j \gamma]$, and that of WLGI and NSIT are  $(\lambda_{th}^{WLGI}$, $1-j \gamma]$ and  $(\lambda_{th}^{NSIT}$, $1-j \gamma]$ respectively.

\subsection{Trade-off between biasedness parameter and unsharpness parameter with respect to the QM violations of LGI, WLGI, NSIT:}

\begin{figure}[!t]
			\resizebox{8.5cm}{6.5cm}{\includegraphics{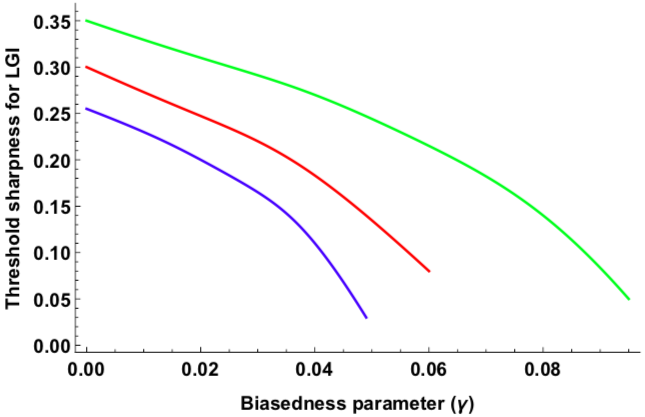}}
			\caption{\footnotesize Green, red and blue line represents the plots of \textbf{the threshold sharpness for LGI $\lambda_{th}^{LGI}$ (the minimum value of the sharpness parameter $\lambda$, above which the QM violation of LGI persists) versus the values of the biasedness parameter ($\gamma$)} associated with the measurements for $j=10$, $j=15$ and $j=20$ respectively, where $j$ represents the spin of the system.} \label{fig11}
		\end{figure}

\begin{figure}[!t]
			\resizebox{8.5cm}{6.5cm}{\includegraphics{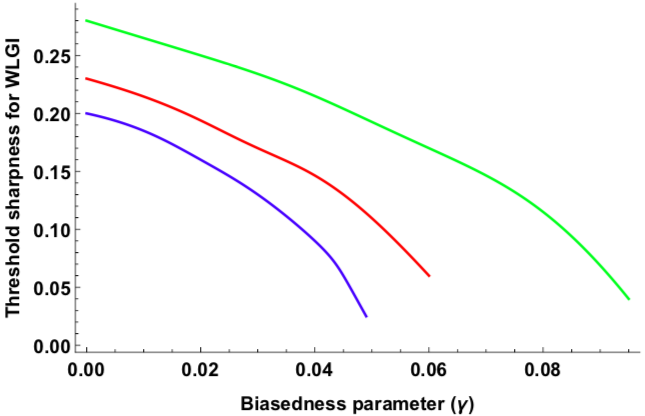}}
			\caption{\footnotesize Green, red and blue line represents the plots of \textbf{the threshold sharpness for WLGI $\lambda_{th}^{WLGI}$ (the minimum value of the sharpness parameter $\lambda$, above which the QM violation of WLGI persists) versus the values of the biasedness parameter ($\gamma$)} associated with the measurements for $j=10$, $j=15$ and $j=20$ respectively, where $j$ represents the spin of the system.} \label{fig21}
		\end{figure}

It is observed that if one introduces the biasedness parameter in an unsharp measurement, then the QM violation of LGI or WLGI persist for a \textit{larger} amount of unsharpness of the measurement in comparison with unbiased case (i. e., $\gamma =0$). The minimum value of the sharpness parameter $\lambda$, above which the QM violations of LGI or WLGI persist, \textit{decreases} with increasing values of the biasedness parameter. This result reflects the trade-off between biasedness and unsharpness of a measurement. On the other hand, introduction of biasedness parameter in the case of violation of NSIT is not interesting as far as $\lambda_{th}^{NSIT}$ is concerned, as it was already minimum ($\lambda_{th}^{NSIT}=0$) for unbiased measurement. These results are shown in Table \ref{tab1}.  In Figures \ref{fig11} and \ref{fig21} we show how the minimum value of the sharpness parameter $\lambda$, above which the QM violations of LGI, WLGI persist, decreases with increasing values of the biasedness parameter $\gamma$ for $j=10$, $j=15$ and $j=20$. For other values of spin ($j$), the nature of the above plots remains unchanged. Since the minimum value of the sharpness parameter $\lambda$, above which the QM violation of NSIT persists, remains the same with increasing values of the biasedness parameter $\gamma$, we have not shown the above plot in the context of QM violation of NSIT.

\begin{figure}[!t]
			\resizebox{8.5cm}{6.5cm}{\includegraphics{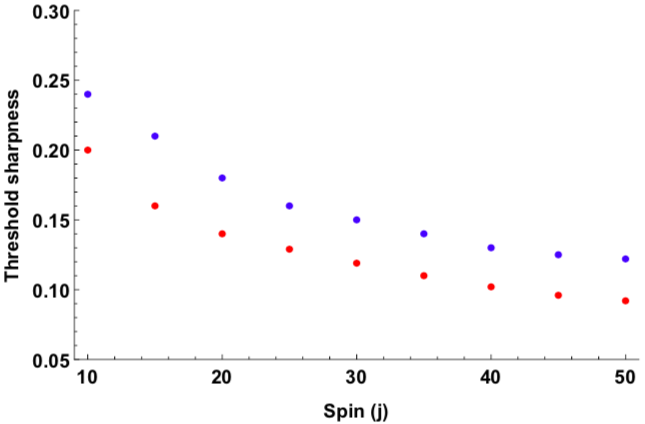}}
			\caption{\footnotesize Blue and red dots represents \textbf{the threshold sharpness of LGI $\lambda_{th}^{LGI}$ and the threshold sharpness of WLGI $\lambda_{th}^{WLGI}$ for different values of  spin $j$ of the system}. In all these cases, the value of the biasedness parameter is $\gamma  = \frac{1}{2j}$.} \label{fig31}
		\end{figure}

We have also evaluated the $\lambda_{th}^{LGI}$, $\lambda_{th}^{WLGI}$ and $\lambda_{th}^{NSIT}$ for a given $j$ at $\gamma$ = mid value in the allowed range of $\gamma$: $0 \leq \gamma < \frac{1}{j}$, i.e., at $\gamma = \gamma_{mid} = \frac{1}{2j}$. In this case we find that for a fixed amount of biasedness incorporated in a measurement (e.g. for $\gamma = \frac{1}{2 j}$), the minimum amount of sharpness parameter necessary for demonstrating QM violations of LGI or WLGI decreases with increasing values of the spin $j$ of the system under consideration. This implies the maximum amount of unsharpness of a measurement (reflecting imprecision of measurement), below which the QM violation of LGI or WLGI persists, increases with the spin value of the system. Again $\lambda_{th}^{NSIT}$ is zero for any values of `j' and $\gamma$. This is consistent with our earlies results \cite{largelgi'2016}. These results are shown in Table \ref{tab2}. In Figure \ref{fig31} we have shown how $\lambda_{th}^{LGI}$ and $\lambda_{th}^{WLGI}$ change with increasing values of spin $j$ (taking $\gamma  = \frac{1}{2j}$).

\begin{figure}[!t]
			\resizebox{8.5cm}{6.5cm}{\includegraphics{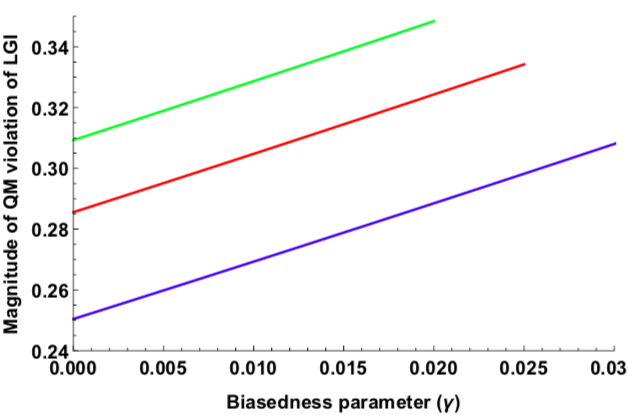}}
			\caption{\footnotesize Green, red and blue line represents the plots of \textbf{the magnitudes of the QM violations of LGI versus the values of the biasedness parameter ($\gamma$)} associated with the measurements for $j=25$, $j=20$ and $j=15$ respectively, where $j$ represents the spin of the system. In all these cases, the value of the sharpness parameter is $\lambda=0.5$.} \label{fig1}
		\end{figure}

\begin{figure}[!ht]
			\resizebox{8.5cm}{6.5cm}{\includegraphics{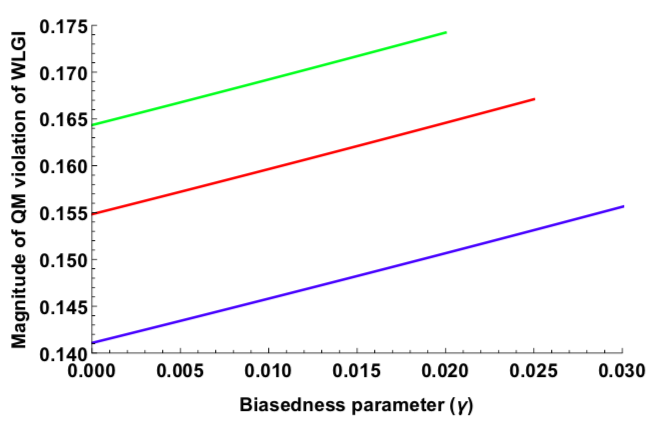}}
			\caption{\footnotesize Green, red and blue line represents the plots of \textbf{the magnitudes of the QM violations of WLGI versus the values of the biasedness parameter ($\gamma$)} associated with the measurements for $j=25$, $j=20$ and $j=15$ respectively, where $j$ represents the spin of the system. In all these cases, the value of the sharpness parameter is $\lambda=0.5$.} \label{fig2}
		\end{figure}

\begin{figure}[!ht]
			\resizebox{8.5cm}{6.5cm}{\includegraphics{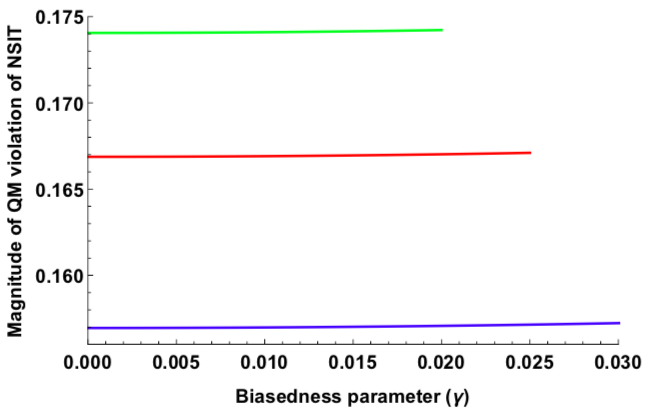}}
			\caption{\footnotesize Green, red and blue line represents the plots of \textbf{the magnitudes of the QM violations of NSIT versus the values of the biasedness parameter ($\gamma$)} associated with the measurements for $j=25$, $j=20$ and $j=15$ respectively, where $j$ represents the spin of the system. In all these cases, the value of the sharpness parameter is $\lambda=0.5$.} \label{fig3}
		\end{figure}

For a given spin $j$ system and for a fixed value of the sharpness parameter $\lambda$, magnitudes of QM violations of all the necessary conditions of MR increase with increasing biasedness introduced in the measurement (i.e. with increasing values of $\gamma$). In contrast to the fact that unsharpness of measurements decreases the magnitudes of QM violations of all the necessary conditions of MR \cite{Saha,Mal-Majumdar,largelgi'2016}, it is clear from the above results that biasedness has a role of facilitating observation of quantum features. In other words, biasedness counters the effect of unsharpness of measurements in demonstrating QM violations of MR. This is illustrated in Table \ref{tab3} and in Figures \ref{fig1}, \ref{fig2} and \ref{fig3} for $\lambda =0.5$. For other values of sharpness parameter $\lambda$, the magnitudes of QM violations of different necessary conditions of MR increase in similar ways with increasing values of $\gamma$. In Figure \ref{fig3} it is not very clear that the QM violation of NSIT condition increases with increasing values of $\gamma$, as the slope is very small. This fact is demonstrated in Table \ref{tab3}.

{\centering
 	\begin{table}
 		\begin{tabular}{|*{4}{c|}}
 			\hline
 			\textit{\textbf{\textbf{}}} & \multicolumn{3}{|c|}{\textit{\textbf{Threshold Sharpness ($\lambda_{th}$) for}}}\\
 			\cline{2-4} 
 			\textit{\textbf{\textbf{}}} & \textit{\textbf{LGI}} & \textit{\textbf{WLGI}} & \textit{\textbf{NSIT}} \\
 			\textit{\textbf{\textbf{}}} & \textit{\textbf{when the}} & \textit{\textbf{when the}} & \textit{\textbf{when the}} \\
 			\textit{\textbf{\textbf{Spin}}} & \textit{\textbf{value of the}} & \textit{\textbf{value of the}} & \textit{\textbf{value of the}} \\
 			\textit{\textbf{\textbf{($j$)}}} & \textit{\textbf{biasedness}} & \textit{\textbf{biasedness}} & \textit{\textbf{biasedness}} \\
 			\textit{\textbf{\textbf{}}} & \textit{\textbf{parameter}} & \textit{\textbf{parameter}} & \textit{\textbf{parameter}} \\
 			\textit{\textbf{\textbf{}}} & \textit{\textbf{is $\gamma$ $= \frac{1}{2 j}$}} & \textit{\textbf{is $\gamma$ $= \frac{1}{2 j}$}} & \textit{\textbf{is $\gamma$ $= \frac{1}{2 j}$}} \\
 			\textit{\textbf{\textbf{}}} & \textit{\textbf{}} & \textit{\textbf{}} & \textit{\textbf{}} \\
 		
 			\hline
 			\hline
 			$15$ &  $0.21$ & $0.16$ & $0$ \\
 			$20$ &  $0.18$ & $0.14$ & $0$ \\ 
 			$25$ &  $0.16$ & $0.13$ & $0$ \\
 			\hline
 		\end{tabular}
 		\caption{For a fixed value of the biasedness parameter ($\gamma$), threshold sharpness for LGI ($\lambda_{th}^{LGI}$) and that for WLGI ($\lambda_{th}^{WLGI}$) decreases with increasing values of spin ($j$); on the other hand, for a fixed value of the biasedness parameter ($\gamma$), threshold sharpness for NSIT ($\lambda_{th}^{NSIT}$) remains the same with increasing values of the spin.} \label{tab2}
 	\end{table}
 	}

		{\centering
			\begin{table}
				\begin{tabular}{|*{6}{c|}} 
					\hline
					
					\textit{\textbf{\textbf{Spin}}} & \textit{\textbf{\textbf{Value}}} & \textit{\textbf{\textbf{Value}}} & \multicolumn{3}{|c|}{\textit{\textbf{Magnitude of the}}}\\
					\textit{\textbf{\textbf{($j$)}}} & \textit{\textbf{\textbf{of}}} & \textit{\textbf{\textbf{of}}} & \multicolumn{3}{|c|}{\textit{\textbf{QM violation of}}}\\
					\cline{4-6} 
					\textit{\textbf{\textbf{}}} & \textit{\textbf{\textbf{$\lambda$}}} & \textit{\textbf{\textbf{$\gamma$}}} & \textit{\textbf{LGI}} & \textit{\textbf{WLGI}} & \textit{\textbf{NSIT}} \\

					\hline
					\hline
					     &       & $0$ &  $0.2504$ & $0.1410$ & $0.1569$ \\
					$15$ & $0.5$ & $0.017$ &  $0.2821$ & $0.1490$ & $0.1570$ \\ 
					     &       & $0.033$ &  $0.3139$ & $0.1571$ & $0.1572$ \\
					\hline
					     &       & $0$ &  $0.2855$ & $0.1548$ & $0.1668$ \\
					$20$ & $0.5$ & $0.012$ &  $0.3096$ & $0.1608$ & $0.1669$ \\ 
					     &       & $0.025$ &  $0.3342$ & $0.1671$ & $0.1671$ \\
					\hline
					     &       & $0$ &  $0.3092$ & $0.1643$ & $0.1740$ \\
					$25$ & $0.5$ & $0.010$ &  $0.3268$ & $0.1692$ & $0.1741$ \\ 
					     &       & $0.020$ &  $0.3484$ & $0.1742$ & $0.1742$ \\
					\hline
				\end{tabular}
				\caption{For a given value of the spin ($j$) and any fixed value of the sharpness parameter ($\lambda$), magnitudes of QM violations of all necessary conditions of MR increase with increasing values of the biasedness parameter ($\gamma$).} \label{tab3}
			\end{table}
			}

In practical scenarios, the measurements are not ideal projective. Nonidealness in measurement arises due to imperfections in the measuring apparatus (which lead to inaccuracies in the outcome readings). This imperfections can be captured theoretically by modeling different kinds of POVM. Unsharp (unbiased) measurement \cite{unsharp1,unsharp2,unsharp5} is one such model, where imprecision present in the measurement is characterised by unsharpness associated with the measurement. Biased unsharp measurement is another form of POVM, where measurement imprecisions are  characterised by biasedness along with unsharpness. In case of dichotomic measurement, unsharpness characterises measurement imprecision which is related to overlap between nonorthogonal probe states \cite{mmh}. On the other hand, biasedness associated with dichotomic spin measurement can be linked with the error in alignment of a Stern-Gerlach apparatus or deviation from Gaussian nature of spatial wave-packet of the incident spin-$\frac{1}{2}$ particles \cite{tba}.

Magnitudes of QM violations of different necessary conditions of MR can be considered as a quantification of nonclassicality associated with the system under consideration. Increasing unsharpness in a measurement decreases the magitudes of QM violations of different necessary conditions of MR for a given spin-$j$ system (where $j$ can take arbitrary integer or half-integer value) \cite{Saha,Mal-Majumdar,largelgi'2016}. On the other hand, from Table \ref{tab3} and Figures \ref{fig1}, \ref{fig2} and \ref{fig3} it is evident that increasing biasedness in a measurement increases the magnitudes of QM violations of different necessary conditions of MR for the given spin-$j$ system. Hence, unsharpness and biasedness in a measurement have opposite roles in the context of demonstrating nonclassicality of the given spin-$j$ system through QM violations of different necessary conditions of MR. Moreover, Table \ref{tab1} and Figures \ref{fig11}, \ref{fig21} show that nonclassicality (probed through QM violation of LGI or WLGI) of the spin-$j$ system under consideration can be demonstrated with a larger amount of unsharpness associated with the measurement if biasedness is introduced in the measurement.

To summarize, nonidealness associated with biasedness can counter the effect of nonidealness associated with unsharpness in reducing quantum features, for example, QM violations of different necessary conditions of MR.

In the present study we have presented QM violation of MR using a particular form of multi-outcome POVM, where the number of outcomes is equal to the dimension of the spin-system. Here the POVM are constructed in order to demonstrate the consequences when various types of nonidealness are associated with each measurement operator, which is projector in ideal scenario (when $\lambda =1$ and $\gamma = 0$). Since the number of projectors associated with spin-projective measurement is equal to the dimension of the spin system, the number of effect operators (or, equivalently, the number of outcomes) of the constructed POVM in the present study is equal to the dimension of the spin-system under consideration. However, one can study QM violation of MR following the treatment adopted in the present study using different forms of POVM where the number of outcomes are greater than or less than the dimension of the system. Physical significance and advantage of using such POVM in the context of QM violation of MR is an interesting issue meriting further investigation.

\subsection{Asymptotic behaviour}
In the asymptotic limit of spin all the necessary conditions of MR yields violation upto algebraic maxima of the corresponding inequalities \cite{largelgi'2016}. Consequently $\lambda_{th}$s for LGI, WLGI and NSIT are all zero \cite{largelgi'2016}. Hence, introduction of biasedness in this limiting case does not affect the results. On the other hand, in the asymptotic limit of spin, only permissible value of the biasedness parameter $\gamma$ is $0$ as can be seen from Equation (\ref{gamma}). It implies that in the limit $j\rightarrow \infty$, biased unsharp measurement becomes unbiased one.

\section{Conclusions} \label{s6}
Joint measurability of pairs of qubit observables was analysed to study a trade-off between the degrees of joint measurability, sharpness and another quantity called biasedness \cite{Stano,bias2}. Sharpness parameter quantifies the precision of a measurement and biasedness arises when one of the measurement outcomes is favoured over the other whatever be the preparation. The most general dichotomic POVM for two level system is characterized by these two parameters.

In this work we attempted to generalise unsharpness and biasedness measure for higher dimensional system. Then we studied a trade-off between the amount of quantum mechanical (QM) violation of MR, unsharpness and biasedness of the relevant measurements. Here QM violation of MR is probed through three necessary conditions of MR, namely, LGI, WLGI and NSIT.

It was known that QM violation of all these conditions decreases with increasing unsharpness of measurements \cite{Saha,Kofler'2008,largelgi'2016}. In contrast with those results here we have shown that QM violations of all the necessary conditions increase with increasing values of the biasedness parameter $\gamma$ irrespective of value of spin for any value of $\lambda (>\lambda_{th})$. Moreover, the minimum value of the sharpness parameter ($\lambda_{th}$) above which the QM violations of LGI or WLGI persists decreases with increasing values of biasedness parameter for any value of spin of the system under consideration.

One important point to be stressed here is that we have used one particular choice of the measurement time intervals in the present study. This particular choice of the measurement time intervals enables us to carry out all the calculations analytically. Since our aim in the present study is to give an idea of the trade-off between the amount of quantum mechanical (QM) violation of MR, unsharpness and biasedness of the relevant measurements, we have not considered arbitrary choices of the measurement time intervals. We conjecture that the nature of the above trade-off presented in this study remains unchanged for arbitrary choices of the measurement time intervals. 

In practical scenarios, the measurements are not ideal projective. The imprecision in measurements can be modeled by employing unsharpness in the measurements. However, it has been shown that classicality is emerged through the presence of unsharpness in the measurements \cite{Saha,Kofler'2008,largelgi'2016}. The present study indicates that introducing biasedness in an unsharp measurement is helpful in demonstrating nonclassicality of a spin-$j$ system. Relating different types of imprecisions of multi-outcome measurements involved in the real experimental scenarios with unsharpness and biasedness is worth to be studied in future.

	\section{Acknowledgements} \label{s7}
The authors acknowledge fruitful discussion with Prof. Dipankar Home. SM acknowledges discussion with Prof. Paul Busch during his visit at S. N. Bose National Centre for Basic Sciences in 2015 and his comments through private communications. DD acknowledges the financial support from University Grants Commission (UGC), Government of India.

\section{Author contribution statements}

Development and design of the problem, calculations have been done by all the authors. All the authors contributed equally to the preparation of the manuscript. On behalf of all authors, the corresponding author states that there is no conflict of interest.

\end{document}